\documentclass[amsmath,aps,showpacs,a4paper,10pt]{revtex4}

 \usepackage{epsf}
 \usepackage{graphicx}    

\begin{document}

 \newcommand{\be}[1]{\begin{equation}\label{#1}}
 \newcommand{\ee}{\end{equation}}
 \newcommand{\bea}{\begin{eqnarray}}
 \newcommand{\eea}{\end{eqnarray}}
 \def\disp{\displaystyle}

 \def\gsim{ \lower .75ex \hbox{$\sim$} \llap{\raise .27ex \hbox{$>$}} }
 \def\lsim{ \lower .75ex \hbox{$\sim$} \llap{\raise .27ex \hbox{$<$}} }

 \begin{titlepage}

 \begin{flushright}
 arXiv:0906.0828
 \end{flushright}

 \title{\Large \bf Tension in the Recent Type Ia Supernovae Datasets}

 \author{Hao~Wei\,}
 \email[\,email address:\ ]{haowei@bit.edu.cn}
 \affiliation{Department of Physics, Beijing Institute
 of Technology, Beijing 100081, China}

 \begin{abstract}\vspace{1cm}
 \centerline{\bf ABSTRACT}\vspace{2mm}
In the present work, we investigate the tension in the recent
 Type Ia supernovae (SNIa) datasets Constitution and Union. We
 show that they are in tension not only with the observations
 of the cosmic microwave background (CMB) anisotropy and the
 baryon acoustic oscillations (BAO), but also with other SNIa
 datasets such as Davis and SNLS. Then, we find the main
 sources responsible for the tension. Further, we make this
 more robust by employing the method of random truncation.
 Based on the  results of this work, we suggest two truncated
 versions of the Union and Constitution datasets, namely the
 UnionT and ConstitutionT SNIa samples, whose behaviors are
 more regular.
 \end{abstract}


 \maketitle

 \end{titlepage}

 \renewcommand{\baselinestretch}{1.6}


\section{Introduction}\label{sec1}
The current accelerated expansion of our universe (see
 e.g.~\cite{r1}) has been one of the most active fields in
 modern cosmology since its discovery from the observations
 of Type Ia supernovae (SNIa)~\cite{r2}. Later on, the
 observations of the cosmic microwave background (CMB)
 anisotropy~\cite{r3,r4} and the large-scale structure
 (LSS)~\cite{r5} confirmed this discovery. Although there are
 already many observational methods to date, SNIa have been
 proven to be one
 of the most powerful tools to probe this mysterious
 phenomenon. In the passed decade, many SNIa datasets have been
 released, while the number and quality of SNIa have continually
 increased. The most familiar SNIa datasets include, for
 instance, Gold04~\cite{r6}, Gold06~\cite{r7}, SNLS~\cite{r8},
 ESSENCE~\cite{r9}, SDSS~\cite{r34}, Davis~\cite{r10}, and,
 most recently, Union~\cite{r11}, Constitution~\cite{r12}.

In~\cite{r11}, 414 SNIa from some heterogeneous compilations
 have been analyzed with the same analysis procedure. After
 selection cuts, this compilation was reduced to 307 SNIa, which
 have been named the Union dataset~\cite{r11}. There are 250 high
 redshift SNIa ($z>0.2$) and 57 low redshift SNIa ($z\leq 0.2$)
 in the Union dataset~\cite{r11}. Very recently, the CfA3
 sample~\cite{r13} has been added to the 307 SNIa Union dataset
 to form the Constitution dataset~\cite{r12}. Originally, the CfA3
 sample consisted of 185 SNIa, which are all at the fairly low
 redshift, i.e., $z<0.08$. After applying the same Union cuts,
 90 SNIa from CfA3 survive. This increases the low redshift
 sample ($z\leq 0.2$) to 147 SNIa in the resulting Constitution
 dataset~\cite{r12}. The 397 SNIa Constitution
 dataset is the largest published, spectroscopically
 confirmed sample to date.

Soon after the release of the 397 SNIa Constitution dataset,
 some authors have used it to study dark energy. For example,
 the authors of~\cite{r14} smoothly reconstructed the
 deceleration parameter $q(z)$ and the $Om(z)$ diagnostic by
 using the Constitution dataset, and found that the cosmic
 acceleration might be slowing down. This is an unexpected
 result in some sense. The authors of~\cite{r15} compared the
 holographic dark energy model~\cite{r16}, the Ricci dark
 energy model~\cite{r17} and the new agegraphic dark energy
 model~\cite{r18}, by using the Constitution dataset and other
 observational data; they found that the holographic dark
 energy model is more favored. This result is also somewhat
 different from the previous studies. Further, the authors
 of~\cite{r19} found that dark energy seemingly did not exist
 in the past and suddenly emerged at redshift $z\sim 0.331$,
 by fitting to the Constitution dataset alone. Needless to
 say, this is a striking claim. In addition, the authors
 of~\cite{r20} found that the dark energy equation of state
 parameter (EoS) deviates from the cosmological constant
 at $z\,\gsim\, 0.5$ significantly, by using the SNIa dataset
 (Union or Constitution) together with the gamma-ray bursts
 (GRBs) data; they noted that the result from Constitution
 dataset is somewhat different from others. In~\cite{r20},
 they also suggested that the deviation might arise from some
 biasing systematic errors in the SNIa and/or GRBs datasets.

These unusual results bring our attention from dark energy
 to the SNIa dataset itself, especially the Constitution SNIa
 dataset. In fact, this situation is reminiscent of the one of
 Gold04 and Gold06 SNIa datasets, which also bring some
 interesting results. In~\cite{r21,r22}, the Gold04 dataset has
 been shown to be in $2\sigma$ tension with the SNLS dataset
 and the WMAP observations. Although the Gold04 sample updated
 to Gold06 several years later, the tension still persisted.
 In~\cite{r23}, the tension and systematics in the Gold06 SNIa
 dataset has been investigated in great detail. In the present
 work, we will follow the method used in~\cite{r23} to study
 the tension in the recent SNIa datasets.


\section{Observational data}\label{sec2}
Before plunging into the issue of the tension in the recent
 SNIa datasets, here we briefly present the cosmological model
 and the observational data. As is well known, the SNIa data
 points are given in terms of the distance modulus
 $\mu_{obs}(z_i)$. The theoretical distance modulus is
 defined as
 \be{eq1}
 \mu_{th}(z_i)\equiv 5\log_{10}D_L(z_i)+\mu_0,
 \ee
 where $\mu_0\equiv 42.38-5\log_{10}h$ and $h$ is the Hubble
 constant $H_0$ in units of $100~{\rm km/s/Mpc}$, whereas
 \be{eq2}
 D_L(z)=(1+z)\int_0^z \frac{d\tilde{z}}{E(\tilde{z};{\bf p})},
 \ee
 in which $E\equiv H/H_0$ and $H$ is the Hubble parameter;
 ${\bf p}$ denotes the model parameters. In the present work,
 we consider the familiar Chevallier-Polarski-Linder (CPL)
 model~\cite{r24}, in which the EoS of dark energy is
 parameterized as
 \be{eq3}
 w_{de}=w_0+w_a(1-a)=w_0+w_a\frac{z}{1+z}\,,
 \ee
 where $w_0$ and $w_a$ are constants. As is well known, the
 corresponding $E(z)$ is given by~\cite{r25,r26,r27}
 \be{eq4}
 E(z)=\left[\Omega_{m0}(1+z)^3
 +\left(1-\Omega_{m0}\right)(1+z)^{3(1+w_0+w_a)}
 \exp\left(-\frac{3w_a z}{1+z}\right)\right]^{1/2},
 \ee
 where $\Omega_{m0}$ is the present fractional energy density
 of pressureless matter. The $\chi^2$ from the SNIa data is
 given by
 \be{eq5}
 \chi^2_{\mu}({\bf p})=\sum\limits_{i}
 \frac{\left[\mu_{obs}(z_i)-\mu_{th}(z_i)\right]^2}{\sigma^2(z_i)},
 \ee
 where $\sigma$ is the corresponding $1\sigma$ error. The parameter
 $\mu_0$ is a nuisance parameter but it is independent of the data
 points. One can perform an uniform marginalization over $\mu_0$.
 However, there is an alternative way. Following~\cite{r22,r28,r29},
 the minimization with respect to $\mu_0$ can be made by expanding
 the $\chi^2_{\mu}$ of Eq.~(\ref{eq5}) with respect to $\mu_0$ as
 \be{eq6}
 \chi^2_{\mu}({\bf p})=\tilde{A}-2\mu_0\tilde{B}+\mu_0^2\tilde{C},
 \ee
 where
 $$\tilde{A}({\bf p})=\sum\limits_{i}\frac{\left[\mu_{obs}(z_i)
 -\mu_{th}(z_i;\mu_0=0,{\bf p})\right]^2}{\sigma_{\mu_{obs}}^2(z_i)}\,,$$
 $$\tilde{B}({\bf p})=\sum\limits_{i}\frac{\mu_{obs}(z_i)
 -\mu_{th}(z_i;\mu_0=0,{\bf p})}{\sigma_{\mu_{obs}}^2(z_i)}\,,
 ~~~~~~~~~~
 \tilde{C}=\sum\limits_{i}\frac{1}{\sigma_{\mu_{obs}}^2(z_i)}\,.$$
 Eq.~(\ref{eq6}) has a minimum for
 $\mu_0=\tilde{B}/\tilde{C}$ at
 \be{eq7}
 \tilde{\chi}^2_{\mu}({\bf p})=
 \tilde{A}({\bf p})-\frac{\tilde{B}({\bf p})^2}{\tilde{C}}.
 \ee
 Since $\chi^2_{\mu,\,min}=\tilde{\chi}^2_{\mu,\,min}$
 obviously, we can instead minimize $\tilde{\chi}^2_{\mu}$
 which is independent of $\mu_0$. Note that the above
 summations are over the whole SNIa dataset.

There are some other observational data relevant to this work,
 such as the observations of the cosmic microwave background
 (CMB) anisotropy~\cite{r3,r4} and the large-scale structure
 (LSS)~\cite{r5}. However, using the full data of the CMB and
 the LSS to perform a global fitting consumes a large amount
 of time. As an alternative, one can instead use the shift
 parameter $R$ from the CMB, and the distance parameter $A$ of
 the measurement of the baryon acoustic oscillation (BAO) peak
 in the distribution of SDSS luminous red galaxies. In the
 literature, the shift parameter $R$ and the distance parameter
 $A$ have been used extensively. It is argued that they are
 model-independent~\cite{r30}, whereas $R$ and $A$ contain the
 main information of the observations of the CMB and the BAO,
 respectively. The shift parameter $R$ is
 defined by~\cite{r30,r31}
 \be{eq8}
 R\equiv\Omega_{m0}^{1/2}\int_0^{z_\ast}
 \frac{d\tilde{z}}{E(\tilde{z})},
 \ee
 where the redshift of recombination is $z_\ast=1090$, which
 has been updated in WMAP5~\cite{r4}. The shift parameter $R$
 relates the angular diameter distance to the last scattering
 surface, the comoving size of the sound horizon at $z_\ast$
 and the angular scale of the first acoustic peak in the CMB
 power spectrum of the temperature fluctuations~\cite{r30,r31}.
 The value of $R$ has been updated to $1.710\pm 0.019$ from
 WMAP5~\cite{r4}. The $\chi^2$ from the shift parameter $R$ is
 $\chi^2_R=(R-R_{obs})^2/\sigma_R^2$. On the other hand, the
 distance parameter $A$ is given by
 \be{eq9}
 A\equiv\Omega_{m0}^{1/2}E(z_b)^{-1/3}\left[\frac{1}{z_b}
 \int_0^{z_b}\frac{d\tilde{z}}{E(\tilde{z})}\right]^{2/3},
 \ee
 where $z_b=0.35$. In~\cite{r32}, the value of $A$ has been
 determined to be $0.469\,(n_s/0.98)^{-0.35}\pm 0.017$. Here
 the scalar spectral index $n_s$ is taken to be $0.960$ which
 has been updated from WMAP5~\cite{r4}. The $\chi^2$ from the
 distance parameter $A$ is $\chi^2_A=(A-A_{obs})^2/\sigma_A^2$.

The best-fit model parameters are determined by minimizing the
 corresponding $\chi^2$. As in~\cite{r25,r33}, the $68\%$
 confidence level (C.L.) is determined by
 $\Delta\chi^2\equiv\chi^2-\chi^2_{min}\leq 1.0$, $2.3$ and
 $3.53$ for $n_p=1$, $2$ and $3$, respectively, where $n_p$ is
 the number of free model parameters. Similarly, the $95\%$
 C.L. is determined by
 $\Delta\chi^2\equiv\chi^2-\chi^2_{min}\leq 4.0$, $6.17$ and
 $8.02$ for $n_p=1$, $2$ and $3$, respectively.


 \begin{center}
 \begin{figure}[tbhp]
 \centering
 \includegraphics[width=1.0\textwidth]{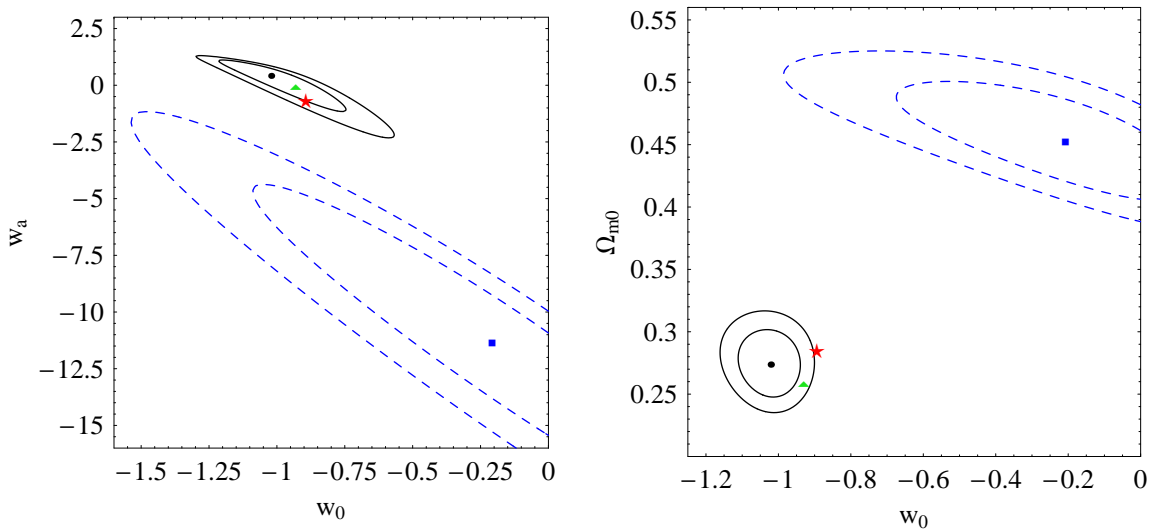}
 \caption{\label{fig1}
 The $68\%$ and $95\%$ C.L. contours in the $w_0-w_a$
 plane and the $w_0-\Omega_{m0}$ plane for the observations of
 SNIa only (blue dashed lines) and SNIa+$A$+$R$ (black solid
 lines). We also show the best-fit values for the observations
 of SNIa only (blue box), SNIa+$A$ (red star), SNIa+$R$ (green
 triangle) and SNIa+$A$+$R$ (black point). These results are
 for the case of Constitution dataset.}
 \end{figure}
 \end{center}


\vspace{-10mm}  


\section{Tension in the recent SNIa datasets}\label{sec3}
At first, we consider the 397 SNIa Constitution dataset. We fit
 the CPL model to the observations of SNIa only, SNIa+$A$,
 SNIa+$R$, and SNIa+$A$+$R$, respectively. The best-fit values
 are presented in Table~\ref{tab1}. In Fig.~\ref{fig1}, we also
 present the $68\%$ and $95\%$ C.L. contours in the $w_0-w_a$
 plane and the $w_0-\Omega_{m0}$ plane. From Fig.~\ref{fig1},
 we find that the Constitution SNIa dataset is in tension
 (significantly beyond $2\sigma$) with the observations of
 the CMB and the BAO.

\begin{center}
 \begin{figure}[tbhp]
 \centering
 \includegraphics[width=1.0\textwidth]{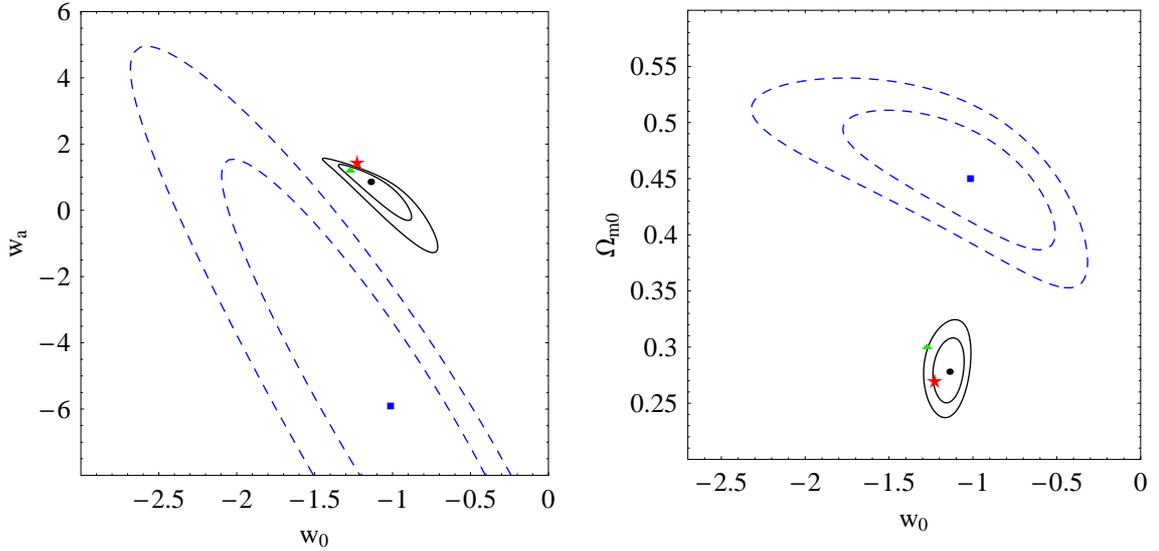}
 \caption{\label{fig2} The same as in Fig.~\ref{fig1}, except
 for the case of Union dataset.}
 \end{figure}
 \end{center}

Next, we turn to the 307 SNIa Union dataset. Similarly, we
 present the results in Table~\ref{tab2} and Fig.~\ref{fig2}.
 From Fig.~\ref{fig2}, we can see that the Union dataset is
 also in tension (beyond $2\sigma$) with the observations of
 the CMB and the BAO, although the corresponding tension is
 weaker than the one in the case of Constitution dataset.

 \begin{table}[tbhp]
 \begin{center}
\vspace{5mm}  
 \begin{tabular}{l|c|c|c|c} \hline\hline
 ~~ Observation & $\chi^2_{min}$ & $\Omega_{m0}$ & $w_0$
 & $w_a$ \\ \hline
 ~~ SNIa & ~~~~ 461.254 ~~~~ & ~~~~ 0.453 ~~~~
 & ~~~~ $-0.207$ ~~~~ & ~~~~ $-11.316$ ~~ \\
 ~~ SNIa+$A$ & 465.438 & 0.286 & $-0.894$ & $-0.628$ \\
 ~~ SNIa+$R$ & 465.739 & 0.260 & $-0.930$ & 0.014 \\
 ~~ SNIa+$A$+$R$ ~~~ & 466.1 & 0.274 & $-1.021$ & 0.413 \\
 \hline\hline
 \end{tabular}
 \end{center}
 \caption{\label{tab1} The $\chi^2_{min}$ and the best-fit
 values of $\Omega_{m0}$, $w_0$ and $w_a$ for the various
 observations. These results are for the case of
 Constitution dataset.}
 \end{table}

 \begin{table}[tbhp]
 \begin{center}
\vspace{6mm}  
 \begin{tabular}{l|c|c|c|c} \hline\hline
 ~~ Observation & $\chi^2_{min}$ & $\Omega_{m0}$ & $w_0$
 & $w_a$ \\ \hline
 ~~ SNIa & ~~~~ 310.091 ~~~~ & ~~~~ 0.451 ~~~~
 & ~~~~ $-1.013$ ~~~~ & ~~~~ $-5.898$ ~~ \\
 ~~ SNIa+$A$ & 310.915 & 0.271 & $-1.228$ & 1.502 \\
 ~~ SNIa+$R$ & 310.911 & 0.302 & $-1.271$ & 1.269 \\
 ~~ SNIa+$A$+$R$ ~~~ & 311.154 & 0.278 & $-1.140$ & 0.859 \\
 \hline\hline
 \end{tabular}
 \end{center}
 \caption{\label{tab2} The same as in Table~\ref{tab1}, except
 for the case of Union dataset.}
 \end{table}

Comparing the cases of the Constitution and Union datasets, we find
 that they have two common features: (i) the best-fit value of
 $\Omega_{m0}$ from SNIa data only is fairly larger than
 $0.3$; (ii) the best-fit value of $w_a$ from SNIa data only
 is strongly negative, namely $w_a$ is much smaller than $-1$.
 On the other hand, they have also two differences: (i) the
 $\chi^2_{min}\sim 460$ is fairly larger than the corresponding
 degrees of freedom $dof\sim 397$ for the Constitution dataset,
 whereas the $\chi^2_{min}\sim 310$ is approximately equal to
 the corresponding $dof\sim 307$ for the Union dataset;
 (ii) for the case of Union dataset, the best-fit value of
 $w_0$ remains about $-1$ for all observations, whereas it is
 not for the case of Constitution dataset.

\begin{center}
 \begin{figure}[tbhp]
 \centering
 \includegraphics[width=1.0\textwidth]{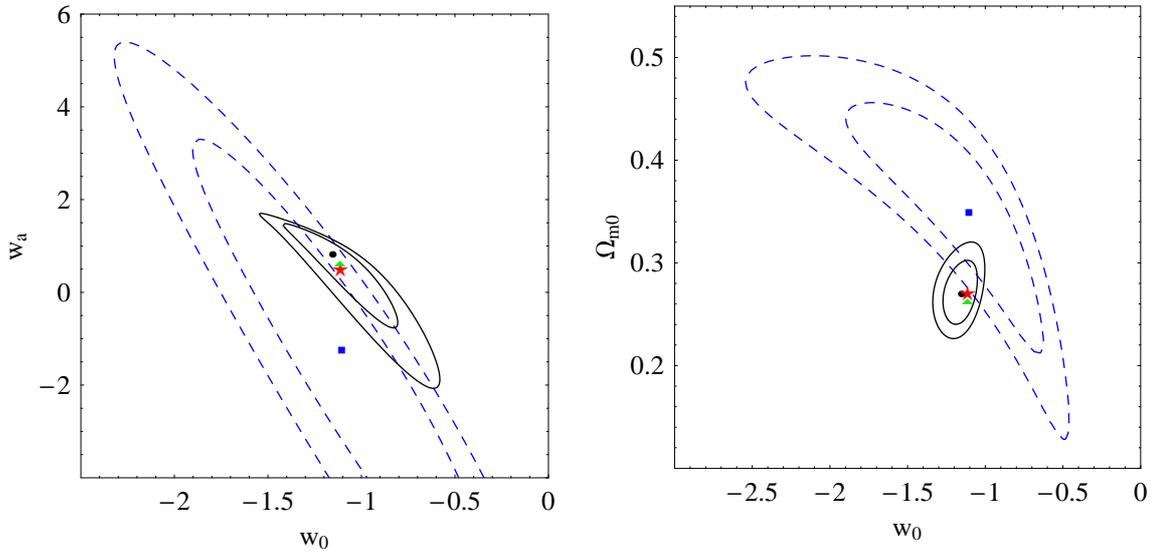}
 \caption{\label{fig3} The same as in Fig.~\ref{fig1}, except
 for the case of Davis dataset.}
 \end{figure}
 \end{center}

Now, one might ask whether or not all SNIa datasets are in
 tension with the observations of CMB and BAO. A third recent
 SNIa dataset is the Davis sample~\cite{r10}, which consists
 of 192 SNIa. We present the corresponding results in
 Table~\ref{tab3} and Fig.~\ref{fig3}. From Fig.~\ref{fig3},
 we see that the Davis SNIa dataset is fully consistent with
 the observations of CMB and BAO. There is {\em no} tension
 in the case of Davis dataset. For the case of 192 SNIa Davis
 dataset, the corresponding $\chi^2_{min}/dof\sim 1$, and its
 $w_0$ remains about $-1$ for all observations, similar to
 the case of Union dataset. However, its best-fit value of
 $\Omega_{m0}$ from SNIa data only is much closer to $0.3$
 than the cases of Union and Constitution datasets.

 \begin{table}[tbhp]
 \begin{center}
\vspace{6mm}  
 \begin{tabular}{l|c|c|c|c} \hline\hline
 ~~ Observation & $\chi^2_{min}$ & $\Omega_{m0}$ & $w_0$
 & $w_a$ \\ \hline
 ~~ SNIa & ~~~~ 195.343 ~~~~ & ~~~~ 0.349 ~~~~
 & ~~~~ $-1.105$ ~~~~ & ~~~~ $-1.229$ ~~ \\
 ~~ SNIa+$A$ & 195.453 & 0.273 & $-1.113$ & 0.526 \\
 ~~ SNIa+$R$ & 195.463 & 0.264 & $-1.115$ & 0.664 \\
 ~~ SNIa+$A$+$R$ ~~~ & 195.485 & 0.270 & $-1.155$ & 0.818 \\
 \hline\hline
 \end{tabular}
 \end{center}
 \caption{\label{tab3} The same as in Table~\ref{tab1}, except
 for the case of Davis dataset.}
 \end{table}


\vspace{-5mm}  


So, the situation is clear now: not all SNIa
 datasets are in tension with the observations of CMB and BAO.
 Besides the Davis SNIa dataset shown above, it is well known
 that the SNLS SNIa dataset is also fully consistent with the
 observations of CMB and BAO~\cite{r21,r22}. Therefore, the
 recent SNIa datasets Constitution and Union are in tension
 not only with the observations of CMB and BAO, but also with
 the other SNIa datasets such as Davis and SNLS. In fact, they
 are in the similar situation of the Gold04 and Gold06 SNIa
 datasets.

 \begin{table}[tbhp]
 \begin{center}
 \begin{tabular}{c} \hline\hline
 {\bf UnionOut\ subset (21 SNIa)}\\ \hline
 {\rm 1992bs,\ 1995ac,\ 1999bm,\ 1997o,\ 2001hu,\ 1998ba,\
 04Pat,\ 05Red,\ 2002hr,\ 03D4au,}\\
 {\rm 04D3cp,\ 03D1fc,\ 03D4dy,\ 03D1co,\ b010,\ d033,\
 g050,\ g055,\ k430,\ m138,\ m226}\\
 \hline\hline
 {\bf ConstitutionOut\ subset (34 SNIa)}\\ \hline
 {\rm 1992bs,\ 1992bp,\ 1995ac,\ 1999bm,\ 1996t,\ 1997o,\
 1995aq,\ 2001hu,\ 1998ba,\ 04Pat,\ 05Red,}\\
 {\rm 2002hr,\ 03D4au,\ 04D3gt,\ 04D3cp,\ 03D4at,\ 03D1fc,\
 04D3co,\ 03D4dy,\ 04D3oe,\ 04D1ak,}\\
 ~~~{\rm 03D1co,\ b010,\ d033,\ f076,\ g050,\ k430,\ m138,\
 m226,\ sn01cp,\ sn02hd,\ sn03ic,\ sn07ca,\ sn07R}~~~\\
 \hline\hline
 \end{tabular}
 \end{center}
 \caption{\label{tabins} The names of SNIa in the UnionOut
 and ConstitutionOut subsets.}
 \end{table}


\section{Outliers with responsibility for the tension}\label{sec4}
In this section, we elucidate sources of the
 tension in the Constitution and Union SNIa datasets. Since
 the Constitution dataset comes from the Union dataset by adding
 90 low redshift SNIa from the CfA3 sample, and the tension
 exists in both Constitution and Union datasets, we speculate
 that the main sources might be in the Union dataset, and the
 tension in the Constitution dataset is mainly inherited from
 the Union dataset. To verify this speculation, we divide the
 Constitution sample into the high and low redshift groups with
 the same dividing line $z=0.2$ in~\cite{r12,r11}, and repeat
 the procedure in the previous section. We find that the
 results for the high and low redshift groups have only little
 difference, and this supports our speculation mentioned above.
 So, we firstly pay our attention to the Union dataset.

In the case of Gold06 SNIa dataset, there are only five subsets
 (see Table~I of~\cite{r23}). Using the method of subset
 truncation (see Sec.~\ref{sec5} below), the authors of~\cite{r23}
 found that the HZSST subset is the main source of the tension
 in the Gold06 dataset. Then, they isolated six SNIa in the
 HZSST subset which are mostly responsible for the tension. On
 the side of Union dataset, there are 13 subsets in it (see
 Table~3 and Table~11 of~\cite{r11}). Even throwing away the
 first five subsets at low redshift, there are still eight
 subsets to be considered. By careful observation, we find that
 three subsets (Barris, Perlmutter, and Riess1998+HZT) are
 notable, because their RMS listed in Table~3 of~\cite{r11} are
 significantly higher than other subsets. Then we try to
 subtract one of these three subsets in turn, and subtract
 these three subsets together from the full Union dataset, and
 see whether or not the tension can be removed. However, we
 find that it does not work in fact. This leads us to speculate
 that the main sources of the tension in the Union dataset
 might be homogeneously distributed in the whole Union dataset,
 namely, they do not concentrate in a single subset listed in
 Table~3 of~\cite{r11}, unlike the case of Gold06 dataset.

\begin{center}
 \begin{figure}[tbp]
 \centering
 \includegraphics[width=0.48\textwidth]{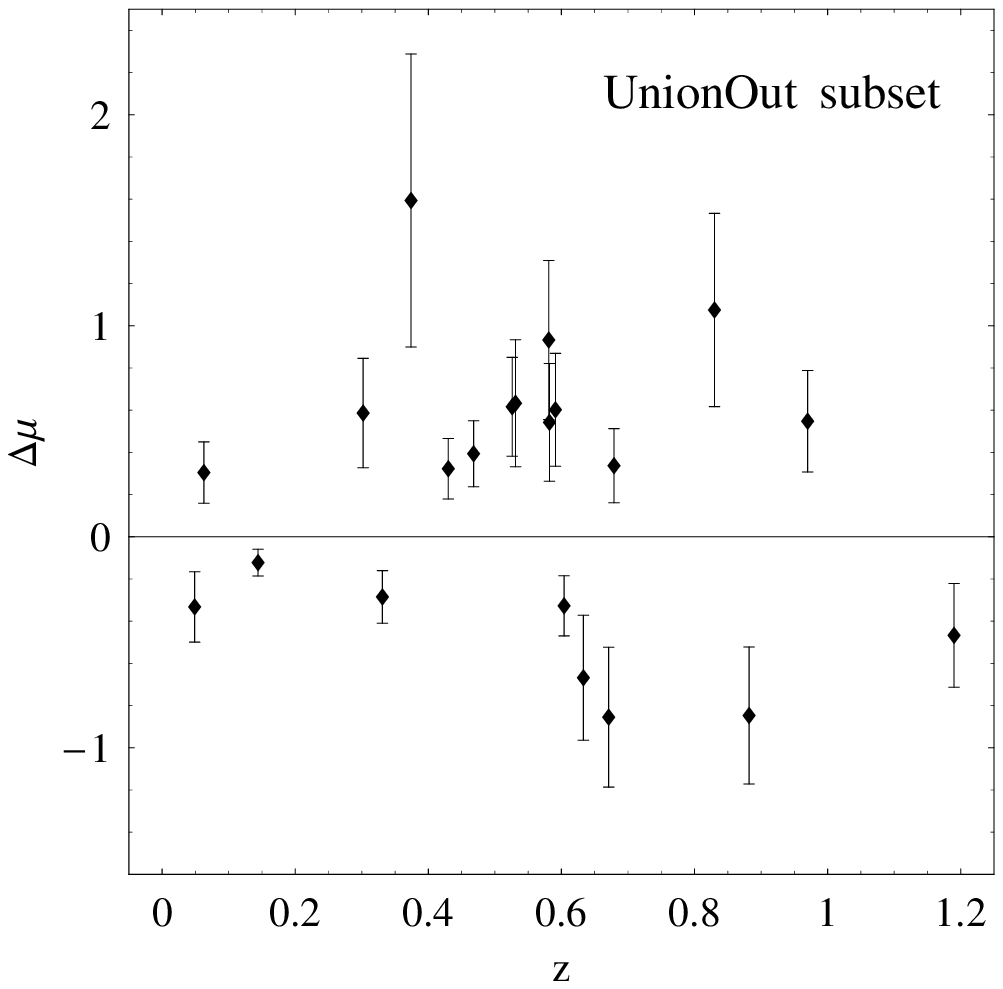}\hfill
 \includegraphics[width=0.48\textwidth]{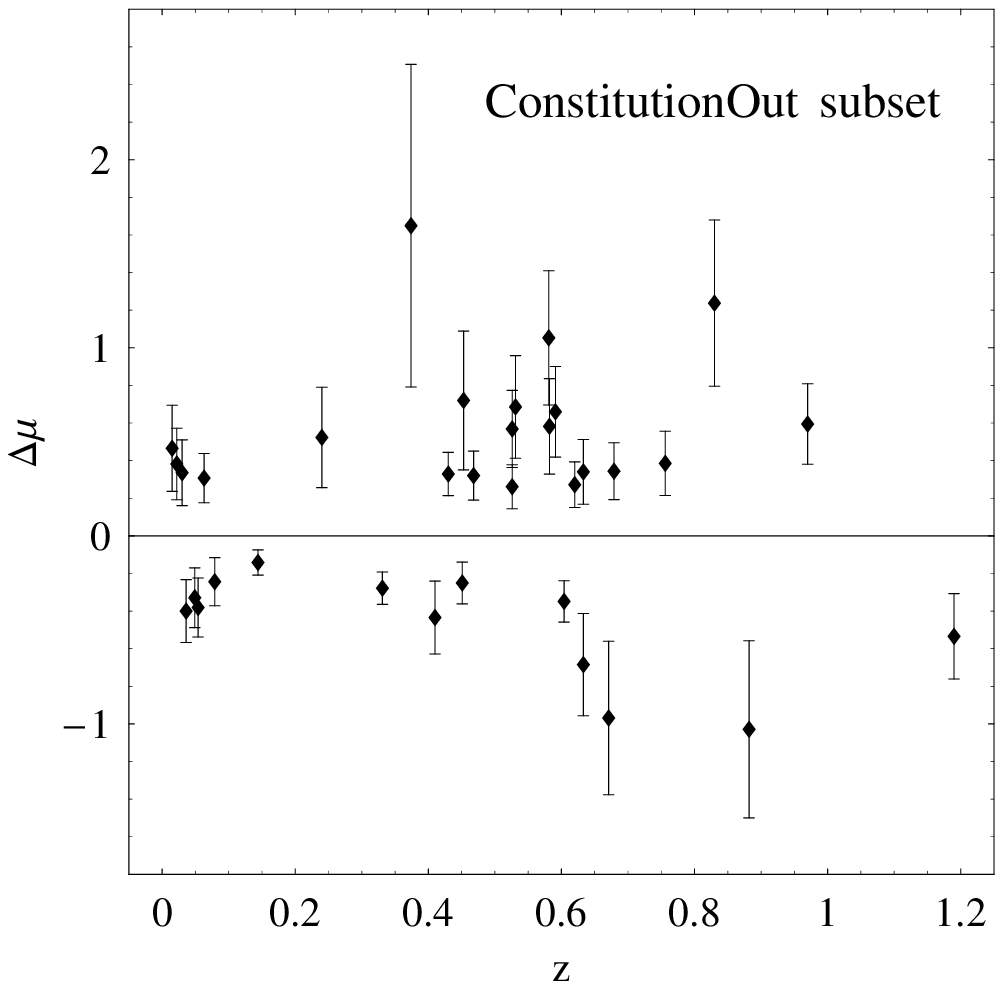}
 \caption{\label{figins}
 $\Delta\mu=\mu_{obs}(z_i)-\mu_{\Lambda\rm CDM}(z_i)$ for
 the SNIa in the UnionOut and ConstitutionOut subsets, where
 the best-fit $\Lambda$CDM are for the whole Union and
 Constitution datasets, respectively. See the text for
 details.}
 \end{figure}
 \end{center}

Here, we follow the simple method used in~\cite{r23} to find
 the outliers responsible for the tension.
 In~\cite{r23}, the distance moduli of the six SNIa which are
 mostly responsible for the tension in Gold06 dataset differ
 by more than $1.8\sigma$ from the $\Lambda$CDM (with
 $\Omega_{m0}=0.28$) prediction. Similarly, we firstly fit
 the flat $\Lambda$CDM model to the whole 307 SNIa in the
 Union dataset, and find that the best-fit parameter is
 $\Omega_{m0}=0.287$ (the corresponding
 $\mu_0=\tilde{B}/\tilde{C}=43.16$). Then, we calculate the
 relative deviation to the best-fit $\Lambda$CDM prediction,
 $|\mu_{obs}-\mu_{\Lambda CDM}|/\sigma_{obs}$, for all the
 307 points. There are 16 SNIa which differ from the best-fit
 $\Lambda$CDM prediction beyond $2\sigma$. However, we find
 that the tension cannot be completely removed by subtracting
 these 16 SNIa from the 307 SNIa Union dataset. We should
 adopt a severer cut. There are 21 SNIa differing from the
 best-fit $\Lambda$CDM prediction beyond $1.9\sigma$, and we
 call them the ``UnionOut'' subset. The names of these 21 SNIa
 are listed in Table~\ref{tabins}. In Fig.~\ref{figins}, we
 also plot their distance modulus deviations relative to the
 best-fit $\Lambda$CDM prediction. These 21 SNIa are indeed
 homogeneously distributed in the whole Union dataset. On the
 other hand, their minimum and maximum redshift are
 $z_{min}=0.0488$ (1995ac) and $z_{max}=1.1900$ (05Red),
 respectively. It is worth noting that if one chooses
 to use instead the XCDM model in which $w_{de}=const.$ to
 find the outliers, the results are almost the same of the
 case of $\Lambda$CDM model used here.

\begin{center}
 \begin{figure}[tbp]
 \centering
 \vspace{1.5mm}  
 \includegraphics[width=1.0\textwidth]{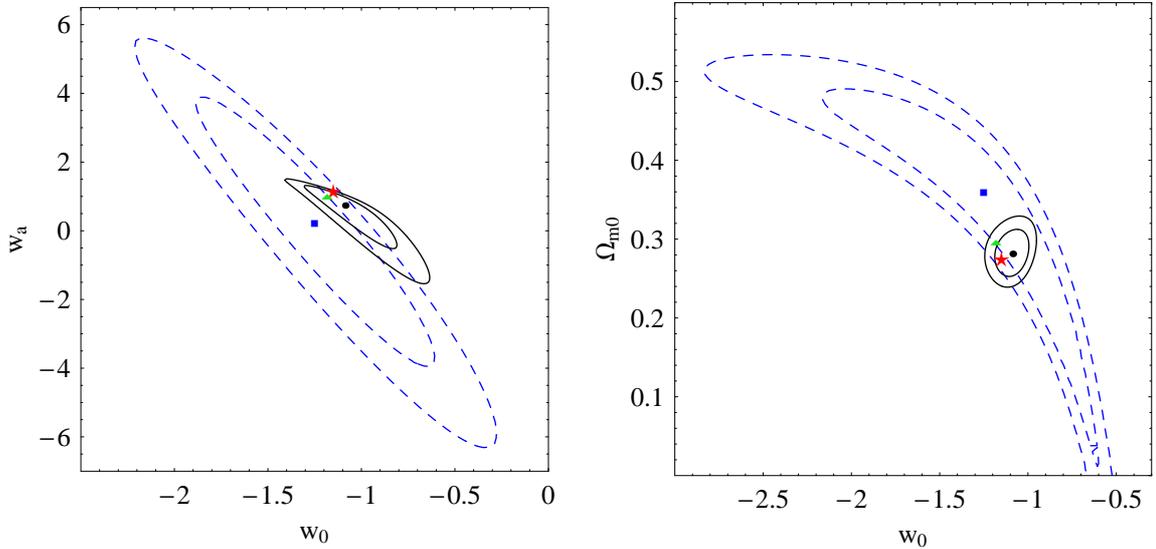}
 \caption{\label{fig4} The same as in Fig.~\ref{fig1}, except
 for the case of the 286 SNIa UnionT sample.}
 \end{figure}
 \end{center}

By subtracting the 21 SNIa UnionOut subset from the whole 307
 SNIa Union dataset, we obtain the so-called ``UnionT'' sample
 (``T'' stands for ``truncated''). Obviously, the UnionT
 sample consists of 286 SNIa. Then, we repeat the same fitting
 as in Sec.~\ref{sec3} for this 286 SNIa UnionT sample, and
 present the corresponding results in Table~\ref{tab4} and
 Fig.~\ref{fig4}. From Fig.~\ref{fig4}, we clearly see that
 the UnionT SNIa sample is fully consistent with the other
 observations, and the tension has been completely removed.
 On the other hand, by dropping the 21 outliers,
 the $\chi^2_{min}$ significantly reduces from
 $\chi^2_{min}\sim 310$ for the 307 SNIa Union dataset (see
 Table~\ref{tab2}) to $\chi^2_{min}\sim 204$ for the 286 SNIa
 UnionT sample (see Table~\ref{tab4}); the corresponding
 $\chi^2_{min}/dof$ has been improved.


 \begin{table}[tbp]
 \begin{center}
 \vspace{5mm}  
 \begin{tabular}{l|c|c|c|c} \hline\hline
 ~~ Observation & $\chi^2_{min}$ & $\Omega_{m0}$ & $w_0$
 & $w_a$ \\ \hline
 ~~ SNIa & ~~~~ 204.060 ~~~~ & ~~~~ 0.360 ~~~~
 & ~~~~ $-1.250$ ~~~~ & ~~~ 0.242 ~~ \\
 ~~ SNIa+$A$ & 204.067 & 0.277 & $-1.151$ & 1.206 \\
 ~~ SNIa+$R$ & 204.067 & 0.298 & $-1.181$ & 1.064 \\
 ~~ SNIa+$A$+$R$ ~~~ & 204.194 & 0.281 & $-1.086$ & 0.735 \\
 \hline\hline
 \end{tabular}
 \end{center}
 \caption{\label{tab4} The same as in Table~\ref{tab1}, except
 for the case of the 286 SNIa UnionT sample.}
 \end{table}

\vspace{-10mm}  

Notice that since dropping these 21 SNIa from the Union
 dataset is enough, we need not adopt other severer
 cuts, such as $1.8\sigma$ or $1.7\sigma$. To preserve the
 number of usable SNIa as much as possible, the cut
 $1.9\sigma$ adopted here is the best choice.

Now, let us turn to the Constitution dataset. Similarly,
 we firstly fit the flat $\Lambda$CDM model to the whole 397
 SNIa in the Constitution dataset, and find that the best-fit
 parameter is $\Omega_{m0}=0.290$ (the corresponding
 $\mu_0=\tilde{B}/\tilde{C}=43.3158$). Then, we calculate the
 relative deviation to the best-fit $\Lambda$CDM prediction,
 $|\mu_{obs}-\mu_{\Lambda CDM}|/\sigma_{obs}$, for all the 397
 points. We adopt the same cut $1.9\sigma$ used in the Union
 dataset. There are 34 SNIa differing from the best-fit
 $\Lambda$CDM prediction beyond $1.9\sigma$, and we call them
 the ``ConstitutionOut'' subset. The names of these 34 SNIa
 are listed in Table~\ref{tabins}. In Fig.~\ref{figins}, we
 also plot their distance modulus deviations relative to the
 best-fit $\Lambda$CDM prediction. These 34 SNIa are indeed
 homogeneously distributed in the whole Constitution dataset.
 On the other hand, their minimum and maximum redshift are
 $z_{min}=0.015$ (sn07ca) and $z_{max}=1.190$ (05Red),
 respectively. Obviously, the UnionOut subset and
 ConstitutionOut subset do heavily overlap. In fact, most SNIa
 of the ConstitutionOut subset come from the Union sample.
 Only the last five SNIa in the ConstitutionOut subset come
 from the CfA3 sample. This confirms our speculation in the
 beginning of the present section that the tension in the
 Constitution dataset is mainly inherited from the Union
 dataset. In other words, the tension is not caused by the
 low redshift CfA3 sample.


\vspace{-10mm}  

\begin{center}
 \begin{figure}[tbp]
 \centering
 \includegraphics[width=1.0\textwidth]{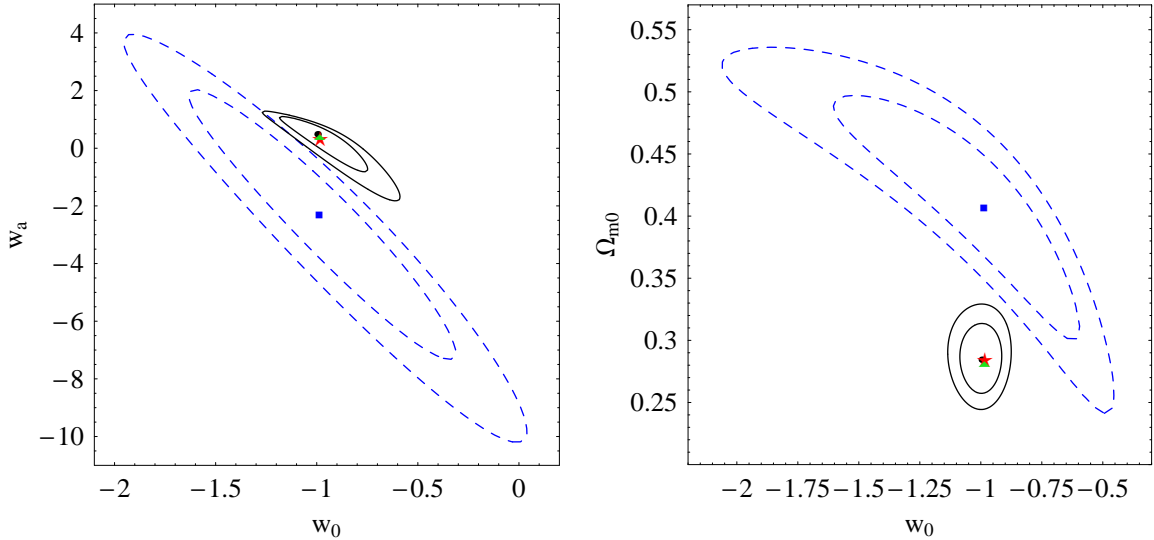}
 \caption{\label{fig5} The same as in Fig.~\ref{fig1}, except
 for the case of the 363 SNIa ConstitutionT sample.}
 \end{figure}
 \end{center}

By subtracting the 34 SNIa ConstitutionOut subset from the
 whole 397 SNIa Constitution dataset, we obtain the so-called
 ``ConstitutionT'' sample. Obviously, the ConstitutionT
 sample consists of 363 SNIa. Then, we repeat the same fitting
 as in Sec.~\ref{sec3} for this 363 SNIa ConstitutionT sample,
 and present the corresponding results in Table~\ref{tab5} and
 Fig.~\ref{fig5}. From Fig.~\ref{fig5}, we see that the
 ConstitutionT SNIa sample is consistent with the other
 observations at the $2\sigma$ level, and the tension has been
 significantly alleviated, especially comparing Fig.~\ref{fig5}
 with Fig.~\ref{fig1}. On the other hand, by dropping the 34
 outliers, the $\chi^2_{min}$ significantly reduces from
 $\chi^2_{min}\sim 461$ for the 397 SNIa Constitution dataset
 (see Table~\ref{tab1}) to $\chi^2_{min}\sim 269$ for the 363
 SNIa ConstitutionT sample (see Table~\ref{tab5}); the
 corresponding $\chi^2_{min}/dof$ has been significantly
 improved.

Notice that if one instead adopt a severer cut $1.8\sigma$
 for the case of Constitution dataset, the results are close
 to the one of $1.9\sigma$ in fact. So, to preserve the number
 of usable SNIa as much as possible, the cut $1.9\sigma$
 adopted here is appropriate.

 \begin{table}[hbp]
 \begin{center}
 \vspace{4mm}  
 \begin{tabular}{l|c|c|c|c} \hline\hline
 ~~ Observation & $\chi^2_{min}$ & $\Omega_{m0}$ & $w_0$
 & $w_a$ \\ \hline
 ~~ SNIa & ~~~~ 268.9 ~~~~ & ~~~~ 0.407 ~~~~
 & ~~~~ $-0.988$ ~~~~ & ~~~ $-2.270$ ~~ \\
 ~~ SNIa+$A$ & 269.13 & 0.286 & $-0.983$ & 0.383 \\
 ~~ SNIa+$R$ & 269.134 & 0.282 & $-0.984$ & 0.438 \\
 ~~ SNIa+$A$+$R$ ~~~ & 269.138 & 0.284 & $-0.996$ & 0.486 \\
 \hline\hline
 \end{tabular}
 \end{center}
 \caption{\label{tab5} The same as in Table~\ref{tab1}, except
 for the case of the 363 SNIa ConstitutionT sample.}
 \end{table}


\vspace{-15mm}  


\section{Verifying the outliers with the method of random truncation}
 \label{sec5}
In the previous section, we have identified the outliers which
 are mostly responsible for the tension in the Union and
 Constitution datasets. Here, we will make them more robust by
 employing the method of random truncation used in~\cite{r23}.

 \begin{table}[bp]
 \begin{center}
 \vspace{2.5mm}  
 \begin{tabular}{ccc} \hline\hline
 $q$ & $q^r$ & $\frac{q-\bar{q}^r}{\sigma_{q^r}}$ \\ \hline
 ~~ $\Omega_{m0}=0.360$ ~~ & ~~~ $\Omega_{m0}^r=0.448\pm 0.025$ ~~~
 & ~~ $-3.6\sigma$ ~~ \\
 $w_0=-1.250$ & $w_0^r=-1.028\pm 0.153$ & $-1.4\sigma$ \\
 $w_a=0.242$ & $w_a^r=-5.715\pm 1.902$ & $+3.1\sigma$ \\
 \hline\hline
 \end{tabular}
 \end{center}
 \caption{\label{tab6} The first column is the best-fit values
 of the quantity $q$ for the 286 UnionT SNIa sample, which can
 be read from Table~\ref{tab4}. The second column is the mean
 best-fit value and corresponding $1\sigma$ range of the
 quantity $q$ for the $N=500$ random truncations. The third
 column is the relative deviation. See the text for details.}
 \end{table}

\begin{center}
 \begin{figure}[tbp]
 \centering
 \includegraphics[width=1.0\textwidth]{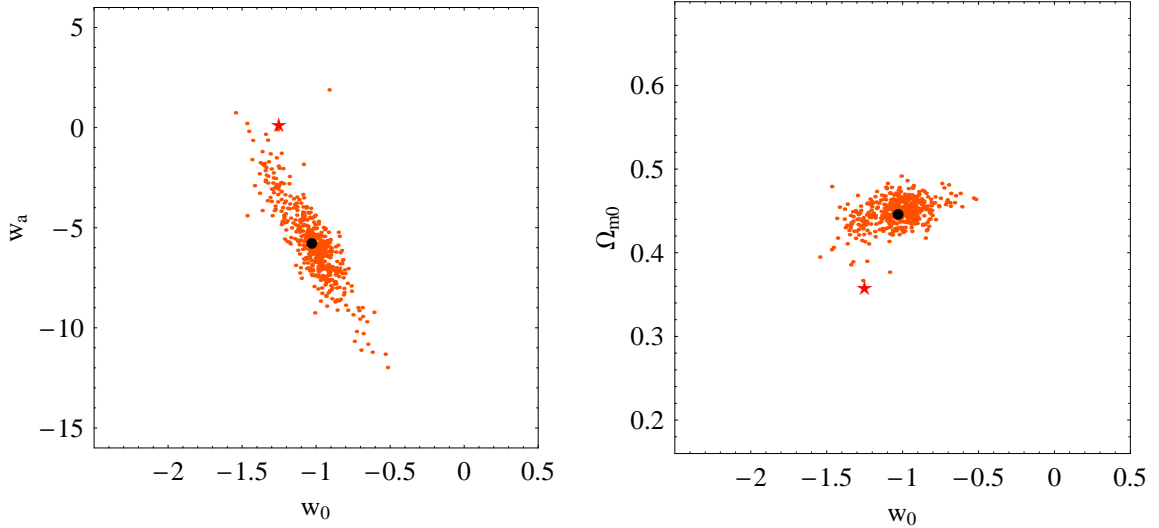}
 \caption{\label{fig6} Comparing the best-fit parameters of
 the 286 UnionT SNIa sample (red stars) with the ones of the
 500 random truncations (orange points) in the $w_0-w_a$ plane
 and the $w_0-\Omega_{m0}$ plane. The mean values of these
 500 points are also indicated by the black solid points. See
 the text for details.}
 \end{figure}
 \end{center}

First, we consider the Union SNIa dataset.
 Following~\cite{r23}, we compare the best-fit values of
 $\Omega_{m0}$, $w_0$ and $w_a$ for the 286 SNIa UnionT data
 with a large number $N$ of corresponding random truncations
 of the Union SNIa data (here we adopt the same number $N=500$
 as in~\cite{r23}). The random truncations involve random
 subtractions of the same number of SNIa and in the same
 redshift range as the UnionOut subset from the full Union
 dataset (notice that $\rm UnionT=Union-UnionOut$). We can
 easily obtain the mean best-fit value $\bar{q}^r$ and the
 $1\sigma$ range $\sigma_{q^r}$ of the quantity
 $q=\Omega_{m0}$, $w_0$ and $w_a$ for these random truncations.
 According to~\cite{r23}, if the best-fit quantity $q$ of the
 UnionT sample is within the $1\sigma$ range of the mean
 best-fit value $\bar{q}^r$ of the random truncations, the
 UnionOut subset is a typical truncation representative of
 the Union dataset and statistically consistent with it. If on
 the other hand the best-fit quantity $q$ of the UnionT sample
 differs from the mean best-fit value $\bar{q}^r$ of the
 random truncations beyond $2\sigma$, we can conclude that
 the UnionOut subset is not a typical truncation and is
 systematically different from the full Union
 dataset~\cite{r23}. We present the results in Table~\ref{tab6}
 and Fig.~\ref{fig6}. Obviously, the UnionOut subset are
 systematically different from the full Union dataset, and its
 21 SNIa are indeed the outliers. Therefore, it is reasonable
 to cut these 21 SNIa outliers in the UnionOut subset from the
 full Union dataset.

\begin{center}
 \begin{figure}[tbp]
 \centering
 \includegraphics[width=1.0\textwidth]{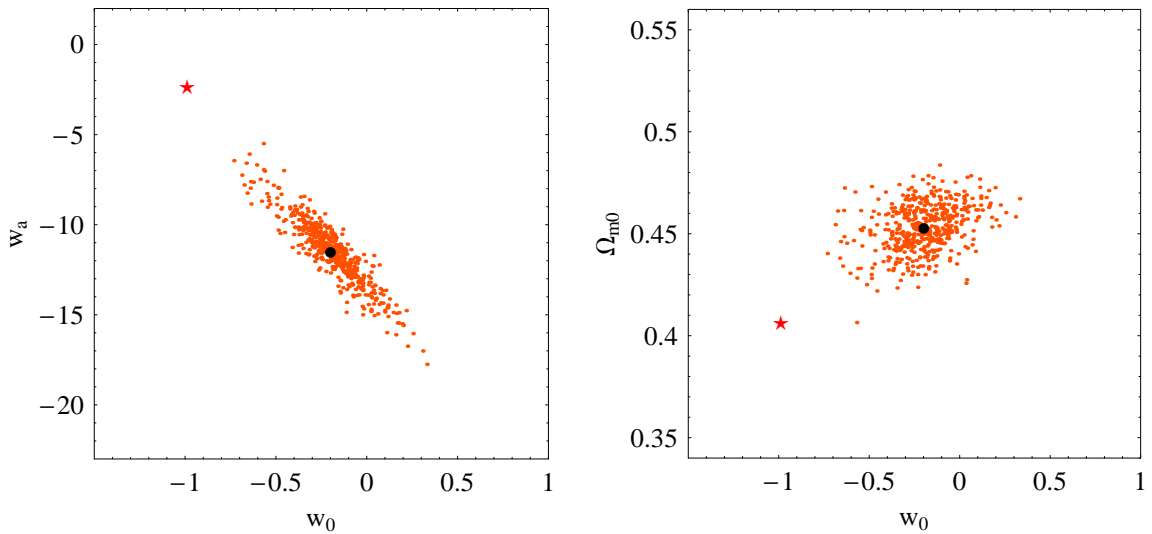}
 \caption{\label{fig7} Comparing the best-fit parameters of
 the 363 ConstitutionT SNIa sample (red stars) with the ones of the
 500 random truncations (orange points) in the $w_0-w_a$ plane
 and the $w_0-\Omega_{m0}$ plane. The mean values of these
 500 points are also indicated by the black solid points. See
 the text for details.}
 \end{figure}
 \end{center}

Next, we turn to the case of the Constitution dataset. Similarly,
 we follow the procedure described above (notice that
 $\rm ConstitutionT=Constitution-ConstitutionOut$), and present
 the results in Table~\ref{tab7} and Fig.~\ref{fig7}.
 Significantly, the ConstitutionOut subset are systematically
 different from the full Constitution dataset, and its 34 SNIa are
 indeed the outliers. Therefore, it is reasonable to cut these
 34 SNIa outliers in the ConstitutionOut subset from the full
 Constitution dataset.


\vspace{-11mm}  


\section{Concluding remarks}\label{sec6}
In the present work, we investigated the tension in the recent
 SNIa datasets Constitution and Union. We have shown that
 they are in tension not only with the observations of CMB
 and BAO, but also with other SNIa datasets such as Davis
 and SNLS. Then, we found the main sources which are
 mostly responsible for the tension. Further, we made this
 more robust by employing the method of random truncation.

Based on the results of this work, we suggest to perform
 the severer selection cut in the Union and Constitution
 SNIa datasets to reject the outliers further. While one uses
 the full 307 SNIa Union dataset and the full 397 SNIa
 Constitution dataset to constrain the cosmological models,
 we strongly recommend use also of the 286 SNIa UnionT sample
 and the 363 SNIa ConstitutionT sample given in this work for
 comparison. In fact, it is anticipated that the results
 might be fairly different, and the unusual features which
 arise from the full 307 SNIa Union dataset and the full 397
 SNIa Constitution dataset might disappear in the cases of
 the 286 SNIa UnionT sample and the 363 SNIa ConstitutionT
 sample.

As communicated by~\cite{r35}, the era of precision cosmology
 has arrived (e.g.~\cite{r34} and \cite{r1}) and there is a
 deemed need for an increased understanding of fundamental
 supernova (SN) physics. In addition, future SN datasets,
 e.g., those from DES and LSST, will rely significantly on
 photometric redshift determinations and SN type
 classifications (particularly in the case of the LSST). The
 details of SN colors will have to be unraveled if purely
 photometric SN are to be successfully used for precision
 cosmology~\cite{r34}. Enlightened by the results of this
 work, one might further consider the possible impact of the
 ConstitutionOut and UnionOut subsets on the understanding
 of fundamental SN physics and SN colors~\cite{r35}. We leave
 this issue as an open question.

 \begin{table}[tbp]
 \begin{center}
 \begin{tabular}{ccc} \hline\hline
 $q$ & $q^r$ & $\frac{q-\bar{q}^r}{\sigma_{q^r}}$ \\ \hline
 ~~ $\Omega_{m0}=0.407$ ~~ & ~~~ $\Omega_{m0}^r=0.453\pm 0.011$ ~~~
 & ~~ $-4.0\sigma$ ~~ \\
 $w_0=-0.988$ & $w_0^r=-0.198\pm 0.171$ & $-4.6\sigma$ \\
 $w_a=-2.270$ & $w_a^r=-11.470\pm 1.804$ & $+5.1\sigma$ \\
 \hline\hline
 \end{tabular}
 \end{center}
 \caption{\label{tab7} The first column is the best-fit values
 of the quantity $q$ for the 363 ConstitutionT SNIa sample, which
 can be read from Table~\ref{tab5}. The second column is the mean
 best-fit value and corresponding $1\sigma$ range of the
 quantity $q$ for the $N=500$ random truncations. The third
 column is the relative deviation. See the text for details.}
 \end{table}


\vspace{-12mm}  


\section*{ACKNOWLEDGEMENTS}
We thank the anonymous referee for expert comments and quite
 useful suggestions, which help us to improve this work. We
 are grateful to Professors Rong-Gen~Cai and Shuang-Nan~Zhang
 for helpful discussions. We also thank Minzi~Feng, as well as
 Shi~Qi, Xin~Zhang, Pu-Xun~Wu, Shuang~Wang and Xiao-Dong~Li,
 for kind help and discussions. In particular, this work was
 triggered by the discussions with them during the period of
 KITPC Program ``Connecting Fundamental Physics with Observations''.
 This work was supported in part by the NSFC under Grant
 No.~10905005, the Excellent Young Scholars Research Fund of
 Beijing Institute of Technology, and the Project of Knowledge
 Innovation Program (PKIP) of Chinese Academy of Sciences under
 Grant No.~KJCX2.YW.W10, which financially supported the KITPC
 Programs.

\vspace{7mm}
\noindent{\bf Note added:} After the submission of the present
 paper, the 1st year SDSS-II supernova sample has been
 released~\cite{r34}. It consists of 103 SNIa with redshifts
 $0.04<z<0.42$. Here, we would like to have a primary analysis.
 Note that the 103 SDSS SNIa dataset was given in both the
 frameworks of using SALT2 and MLCS2k2 light curve fitters.
 Similarly, we fit the CPL model to the observations of SNIa
 only, SNIa+$A$, SNIa+$R$, and SNIa+$A$+$R$, respectively. The
 best-fit values are presented in Table~\ref{tab8}. We can see
 that the best-fit parameters $\Omega_{m0}$, $w_0$ and $w_a$
 for the case of SNIa only (both SALT2 and MLCS2k2) are fairly
 unusual. On the other hand, we find that in the case of SNIa
 only, the constraints on the model parameters $\Omega_{m0}$,
 $w_0$ and $w_a$ are very loose. These results might be mainly
 due to the relatively narrow redshift range ($0.04<z<0.42$)
 and the relatively small number of the SDSS SNIa sample.
 Considering these issues, unlike the cases of the Constitution,
 Union and Davis datasets, the tension analysis might be not so
 robust for the case of the SDSS SNIa dataset. However, since
 the SDSS SNIa dataset fills in the redshift ``desert'' between
 low- and high-redshift SN surveys, it deserves further
 attentions in the SN studies.

 \begin{table}[tbhp]
 \begin{center}
 \vspace{5mm}  
 \begin{tabular}{l|c|c|c|c} \hline\hline
 ~~ Observation & $\chi^2_{min}$ & $\Omega_{m0}$ & $w_0$
 & $w_a$ \\ \hline
 ~~ SNIa (SALT2) & ~~~~ 122.305 ~~~~ & ~~~~ 0.999 ~~~~
 & ~~~~ $90.379$ ~~~~ & ~~~~ $-1776.16$ ~~ \\
 ~~ SNIa (SALT2) + $A$ & 128.902 & 0.297 & $-1.081$ & 2.147 \\
 ~~ SNIa (SALT2) + $R$ & 128.909 & 0.406 & $-1.234$ & 1.417 \\
 ~~ SNIa (SALT2) + $A$ + $R$ ~~~ & 128.973 & 0.296 & $-0.881$
 & 0.252 \\
 \hline\hline
 ~~ SNIa (MLCS2k2) & ~~~~ 173.159 ~~~~ & ~~~~ 0.970 ~~~~
 & ~~~~ $29.806$ ~~~~ & ~~~~ $-438.023$ ~~ \\
 ~~ SNIa (MLCS2k2) + $A$ & 181.143 & 0.315 & 1.497 & $-19.822$ \\
 ~~ SNIa (MLCS2k2) + $R$ & 181.646 & 0.137 & 1.053 & $-14.190$ \\
 ~~ SNIa (MLCS2k2) + $A$ + $R$ ~~~ & 194.408 & 0.311 & $-0.532$
 & $-0.877$ \\
 \hline\hline
 \end{tabular}
 \end{center}
 \caption{\label{tab8} The same as in Table~\ref{tab1}, except
 for the case of SDSS SNIa dataset.}
 \end{table}

\vspace{-3mm}  

\renewcommand{\baselinestretch}{1.2}


\end{document}